\documentclass[12pt,preprint]{aastex}

\shortauthors{Okamoto et al.}
\shorttitle{Resonant Absorption in a Solar Prominence - Observations}
\begin{document}

\title{Resonant Absorption of Transverse Oscillations and Associated Heating in a Solar Prominence. I- Observational aspects}
\author{\textsc{
Takenori J. Okamoto,$^{1,8}$\altaffiltext{8}{Current address: STEL, Nagoya University, Aichi 464-8601, Japan}
Patrick Antolin,$^{2}$
Bart De Pontieu,$^{3,4}$
Han Uitenbroek,$^{5}$
Tom Van Doorsselaere,$^{6}$
Takaaki Yokoyama$^{7}$
}}
\affil{
$^{1}$ISAS/JAXA, Sagamihara, Kanagawa 252-5210, Japan\\
$^{2}$National Astronomical Observatory of Japan, Mitaka, Tokyo 181-8588, Japan\\
$^{3}$Lockheed Martin Solar and Astrophysics Laboratory, B/252, 3251 Hanover St., Palo Alto, CA 94304, USA\\
$^{4}$Institute of Theoretical Astrophysics, University of Oslo, P.O. Box 1029 Blindern, N-0315 Oslo, Norway\\
$^{5}$National Solar Observatory, PO Box 62, Sunspot, NM 88349, USA\\
$^{6}$Centre for Mathematical Plasma Astrophysics, Mathematics Department, KU Leuven, Celestijnenlaan 200B bus 2400, B-3001 Leuven, Belgium\\
$^{7}$The University of Tokyo, Hongo, Bunkyo, Tokyo 113-0033, Japan
}
\email{okamoto@solar.isas.jaxa.jp}

\begin{abstract}
Transverse magnetohydrodynamic (MHD) waves have been shown to be ubiquitous in the solar atmosphere and can in principle carry sufficient energy to generate and maintain the Sun's million-degree outer atmosphere or corona. However, direct evidence of the dissipation process of these waves and subsequent heating has not yet been directly observed. Here we report on high spatial, temporal, and spectral resolution observations of a solar prominence that show a compelling signature of so-called resonant absorption, a long hypothesized mechanism to efficiently convert and dissipate transverse wave energy into heat. Aside from coherence in the transverse direction, our observations show telltale phase differences around 180$^{\circ}$ between transverse motions in the plane-of-sky and line-of-sight velocities of the oscillating fine structures or threads, and also suggest significant heating from chromospheric to higher temperatures. Comparison with advanced numerical simulations support a scenario in which transverse oscillations trigger a Kelvin-Helmholtz instability (KHI) at the boundaries of oscillating threads via resonant absorption. This instability leads to numerous thin current sheets in which wave energy is dissipated and plasma is heated. Our results provide direct evidence for wave-related heating in action, one of the candidate coronal heating mechanisms.
\end{abstract}

\keywords{waves --- Sun: chromosphere --- Sun: transition region}

\section{Introduction}

Since 15 years, transverse oscillations have been detected quite regularly in the solar corona. The \emph{TRACE} observations \citep{han99} starting from 1998 showed plenty of flare-induced oscillations \citep{nak99,sch99,asc99}. They were successfully
used for seismology of the solar corona, by estimating the magnetic field \citep{nak01}, the loop radial structure \citep{asc03}, the vertical density scale height \citep{and05} and Alfv\'en transit times \citep{arr07}.

Since 2007, with the advent of the CoMP instrument and the Hinode satellite \citep{kos07}, we are now convinced that the entire solar atmosphere is filled with low-amplitude propagating and standing Alfv\'enic waves \citep{tom07,oka07,dep07}, even in the absence of a flare excitation. Subsequent studies with CoMP \citep{tom09}, Hinode and the Solar Dynamics Observatory \citep[\emph{SDO};][]{pes12} spacecrafts revealed the wave properties and showed their presence in prominences \citep{sch13}, spicules \citep{he09,oka11} and coronal loops \citep{mci11,wan12,nis13,anf13}. These oscillations are the manifestation of magnetohydrodynamic (MHD) waves \citep[e.g.,][and references therein]{oli09,mac10,arr12,mat13}. Such waves may play an important role in the so-called coronal heating problem by supplying the Sun's hot outer atmosphere with energy \citep[e.g.,][]{alf47,par12}. The energy carried by these Alfv\'enic waves is, in principle, strong enough to heat the corona and accelerate the solar wind \citep{uch74,ant10,vanB11,mat14}. However, although such waves are sometimes damped on short spatial and temporal scales in observations \citep{asc99,pas10,pas11}, it is still unclear whether any significant dissipation (and thus heating) occurs in the corona, which has very low resistivity.

The dissipation of transverse MHD waves in an inhomogeneous medium has been studied for decades \citep[e.g.,][]{ion78,hol90,sak91,goo92,van04}. Theoretically it is expected that the transverse, coherent motion is subject to damping by resonant absorption in an inhomogeneous prominence thread \citep{arr08,sol12}.
Resonant absorption converts the transverse wave into azimuthal motions with a fine spatial scale in the tube's boundary layer \citep{ver10,sol10,arr11}. The resonant and amplified azimuthal motions are prone to be unstable to the Kelvin-Helmholtz instability (KHI), because of the large shear motions. The KHI deforms the boundary layer and leads to enhanced dissipation of the wave energy into heat in thin, turbulent current sheets \citep{kar93,ofm94,lap03,ter08b}. Recent simulations by \citet{ant14} show that even small amplitude oscillations can lead to the instability, whose vortices combined with line-of-sight (LOS) effects result in strand-like structures within coronal loops, but at scales that are difficult to resolve with current coronal instruments.

This study is composed of two parts. In this first paper, unique coordinated observations with the Solar Optical Telescope \citep[SOT;][]{tsu08,shi08,sue08} of \emph{Hinode} and NASA's recently launched \emph{Interface Region Imaging Spectrograph} \citep[\emph{IRIS;}][]{dep14} satellite are used to provide evidence for a telltale sign of resonant absorption and associated heating in an active region solar prominence. We extend the three-dimensional MHD model used in \citet{ant14} to a scenario matching our observations. Through radiative transfer, we show that the observed transverse dynamical coherence, the specific phase relations between the azimuthal motions and displacement of the threads and the characteristic heating are a characteristic feature of resonant absorption combined with the KHI. The details of the numerical model and a more thorough analysis of the observational signatures of transverse MHD waves in prominences are provided in \citet{ant15}, hereafter Paper~2.

\section{Observation and Data Reduction}

\emph{IRIS} observed an active region prominence on the southeast limb of the Sun from 09:00 to 11:00 UT on 19 October 2013 (Figure \ref{fig1}). Four-step sparse raster scans, which have 1\arcsec\ gap between neighboring slit positions, were performed together with the slit-jaw images of the \ion{Si}{4} and \ion{Mg}{2} lines. The cadence was 10 s for the slit-jaw images and 20 s for spectra at each position with the spatial and spectral resolutions of 0.33\arcsec --0.40\arcsec\ (240--290 km on the Sun) and 3 km s$^{-1}$. The exposure time was 4 s in both. Coordinated observations with the SOT were performed from 09:09 to 10:27 UT. The SOT obtained \ion{Ca}{2} H-line filtergrams with a cadence of 8 s, exposures of 1.2 s, and the spatial resolution of 0.2\arcsec\ (150 km). The Atmospheric Imaging Assembly \citep[AIA;][]{lem12} aboard \emph{SDO} provided context filtergrams of multiple wavelengths with a cadence of 12 s.

Here we used level-2 data of \emph{IRIS}, which are calibrated by dark current subtraction, flat-field correction, and geometrical correction. Remaining offsets by a few pixels between the slit-jaw images and spectral data were found, and then we shifted the images to align with fiducial marks that appear on the slit-jaw images and the spectra. Next, we performed co-alignment of time series of the slit-jaw images. Large-amplitude orbital variation of \emph{IRIS} pointings, which is called wobble, is typically removed at the moment of planning \citep{dep14}, but small misalignment less than 1\arcsec\ still remained. With cross correlation of two neighboring-time images of \ion{Si}{4}, the offset information was derived and applied for all \emph{IRIS} slit-jaw and spectral data. Hence, one can find slow drifts of the slit locations in the time-series of images (see the online movie A). Although this impact was small enough in a timescale of focused oscillations (10--20 min), we considered it when we chose the nearest locations to investigate the LOS velocity of oscillating threads. Lastly, we performed co-alignments between the \emph{IRIS} slit-jaw images (\ion{Si}{4}) and the SOT filtergrams (\ion{Ca}{2}) by adjusting their plate scales, rotating, and shifting images. To emphasize the faint prominence structures, a radial-density filter was applied for the disk parts (also part of spicules) of all images. No gamma value is applied for these images.

We mainly use \ion{Mg}{2} k (2796 \AA, typical of plasma at 10,000 K) spectra of the prominence to determine their LOS velocity as a function of time. Even though the spectra show some central reversal, we perform centroiding to determine the LOS velocity which provides a good estimate of the velocity \citep{lee13}. Single Gaussian functions are used to fit the \ion{Mg}{2} spectra to determine the centroid LOS velocity. Comparison of these LOS velocity signals with oscillatory displacements of the same threads in the plane-of-sky (POS), obtained in the \ion{Ca}{2} H (3968 \AA, 10,000 K) passband of the SOT, allows to reconstruct the three-dimensional flow generated by the waves.

\section{Methods and Results}

\subsection{Height dependence of threads and signatures of heating}
The \ion{Ca}{2} H-line movie shows that the prominence consists of numerous horizontal threads that move horizontally with various speeds (10--40 km s$^{-1}$). Space-time plots tracking the flow reveal that many of these threads oscillate in the POS, with signatures of damping. These features can also be seen in the \ion{Mg}{2} k and \ion{Si}{4} (1403 \AA, 80,000K) \emph{IRIS} images, despite the coarser spatial resolution. In the space-time plot of LOS velocity derived from \ion{Mg}{2} k spectra along one slit location, we can find oscillatory patterns at all heights (Figure \ref{fig2}(b)). Moreover, the upper region shows higher velocity amplitudes than the lower region. Similarly, the line width of \ion{Mg}{2} k is larger in the upper region than in the lower region (Figure \ref{fig2}(c)). From the histogram of LOS velocity (Figure \ref{fig2}(d)), the higher region (regions 3 and 4, indicated by yellow and red lines, respectively) has more high-speed components with more than 15 km s$^{-1}$, while velocity in the lower region is more confined around 0 km s$^{-1}$. This is consistent with the large-amplitude motions seen in the \ion{Ca}{2} movie. Similarly, the histogram of line width (Figure \ref{fig2}(e)) also shows stronger broadening in the higher region than in the lower region. We note that the line width in Figure \ref{fig2} includes the instrument line broadening (5.5 km s$^{-1}$ in FWHM). The movie shows that the lower region has more LOS superposition with a multitude of threads overlapping. One would expect such superposition to lead to broader spectral line profiles. However, this is not found, suggesting that the difference in non-thermal line broadening between the low and high regions is instead caused by increased turbulence or unresolved wave motions, which can lead to heating.

Further careful investigation reveals that the horizontal threads in \ion{Ca}{2} H have smaller length and shorter lifetime at higher altitudes. As the moving threads in the \ion{Ca}{2} line fade away, co-spatial threads appear in the hotter \ion{Si}{4} line with similar horizontal speeds. Space-time plots of the \ion{Ca}{2} and \ion{Si}{4} images show this more clearly (Figure \ref{fig3}). Even though some locations have coexistence of cooler and hotter materials, trajectories of several threads show the transition from lower to higher temperature. These fade out finally. This indicates that the cool plasma is heated to coronal temperature through the mid-temperature range of the \ion{Si}{4} line.

Two kinds of intensity variability can be observed in the online movie A. First is a variability produced by the dynamic nature of threads linked to flows and waves. This effect, as discussed in the next section, is mostly behind the fact that a fixed slit can only capture parts of the transverse oscillations in the POS, and entails a timescale on the order of one period or less. The second kind of variability corresponds to a gradual intensity change from the chromospheric emission in \ion{Ca}{2}~H to emission in the transition region range of the \ion{Si}{4} line, and can be appreciated in Figure \ref{fig3}. This variability is therefore linked to heating and occurs on slightly longer timescales of $1-2$ periods. Both effects, dynamics and heating, are intertwined and make the exact determination of lifetimes of threads difficult, even more from the strong line-of-sight projection effects (especially in \ion{Ca}{2}~H). In the online movie A we observe threads appearing and disappearing in the top part of the prominence on timescales of $10-15$ min, i.e. a couple of periods of the oscillations. Such lifetimes appear to be longer in the lower part of the prominence, where projection effects are stronger but also where dynamics are reduced, as explained previously.

\subsection{Transverse oscillations}

The \ion{Ca}{2} movie clearly shows the ubiquity of transverse motions of threads. Oscillations over several periods are not easy to spot in this kind of dynamic prominence. This is largely due to the horizontal flow, which makes the oscillations appear non-stationary \citep{oka07}, and move them quickly away from the fixed slit positions. Also, as shown by Figure \ref{fig4} and the \ion{Ca}{2} movie, vertical displacements of large horizontal sections are observed. Both effects complicate the visibility of multiple periods. To clearly indicate the presence of transverse oscillations and better visualize the longer periodic oscillations of the threads, space-time plots at different horizontal locations following the flow in the threads are provided in the online movies C--F. One can clearly find 2- to 3-period oscillations of P1 and P3 threads with damping of their amplitudes (marked by arrows in the movies). Although the abundance of oscillations is also clear for P2 and P4, the large quantity of threads and their very dynamic nature impedes a fully clean tracking of oscillations beyond 2 periods. Signatures of damping are detected, but also constant and even some cases of increasing amplitude oscillations.

\subsection{Characteristic phase differences and dynamical coherence}\label{phasedif}

Next, we focus on four oscillating cool threads that can be seen clearly and individually with little superposition in the \ion{Ca}{2} images (P1--P4 in Figure \ref{fig1}(a)) to investigate the phase relationship between the transverse displacement in the \ion{Ca}{2} line and LOS velocity in the \ion{Mg}{2} line (Figure \ref{fig4}). We show the combined velocity signal from the nearest slit location at each time. Despite the fact that the threads oscillate across the \emph{IRIS} slit locations at all times, the LOS velocity signals detected around the threads are strongly coherent (Figure \ref{fig5}).  Interestingly, the oscillating threads show phase differences around 90$^{\circ}$ and 180$^{\circ}$ between the transverse displacement in the POS and the LOS velocity. Large phase differences above 90$^{\circ}$, in particular 180$^{\circ}$, do not match theories of the classic kink mode with discontinuous boundary in the density profile, nor those invoking classic axisymmetric torsional Alfv\'en waves or rotation \citep{goo14}.

Figure \ref{fig5} shows the locations of the slits around the threads for positions P1--P4. Besides the out-of-phase relation between the POS motion of the thread and the LOS velocity, another striking feature is the existence of coherence in the Doppler signal across a significant transverse distance on the order of a thread diameter or more, away (above and below) the \ion{Ca}{2}~H thread. Such coherence may come as a surprise if one assumes that the external medium dominates our signal -- if that were the case one would expect a loss of coherence. Besides coherence in the LOS velocity, the existence of coherent transverse POS motion can also be clearly seen in the online movies C--F. Such coherent POS motion of thread-like structure seems to be a common characteristic of transverse waves in prominences \citep{oka07,lin09,lin11,hil13} and in loops, as evidenced by siphon flows \citep{ofm08} or coronal rain \citep{ant11}. In the present observations we show evidence for the first time that the coherent motion extends to the azimuthal flow generated by the transverse wave.

It is important to notice that the \ion{Ca}{2}~H threads appear surrounded by \ion{Mg}{2}~k emission, as evidenced by Figure \ref{fig1}. This fact, together with the existence of dynamical coherence in the POS and along the LOS at significant distance across the threads suggests the existence of larger prominence flux tubes containing the threads. This scenario will be further explored in the next section.

\subsection{The transverse MHD wave model}\label{model}

To understand the observational results and in particular the peculiar phase difference, we use three-dimensional MHD simulations of a transversely oscillating prominence flux tube, combined with radiative transfer modeling. The flux tube has initial internal to external electron number density ($n_i/n_e$) and temperature ($T_i/T_e$) ratios of 10 and 1/100, respectively. The transition layer between the internal and external medium has a width $l/R\approx 0.4$, where $R$ is the flux tube radius. For the forward modeling we give typical values found in prominences $n_{i}=10^{10}~$cm$^{-3}$ and $T_i =10^{4}~$K \citep{tan95,via15}. At time $t=0$ the flux tube is subject to a transverse perturbation generating a fundamental standing transverse MHD wave with longitudinal wavenumber $k \approx 0.015/R$, amplitude $v_0 = $8~km~s$^{-1}$ and phase speed of 776~km~s$^{-1}$. The external plasma-$\beta$ is 0.01 and the fully ionized external medium has a magnetic field value of 18.6~G. We assume a long ($L=200$ Mm) and thin ($R=1$ Mm) prominence flux tube, matching the observations. For further details please see Paper~2. 

As expected from theory, resonant absorption in the simulation sets in immediately and the energy from the purely transverse kink mode is rapidly transferred to the azimuthal waves in the resonant layer at the boundary of the tube (the online movie~B). This induced azimuthal component of the velocity is, at first, in phase with the dipole-like azimuthal flow outside the tube, thus exhibiting a 90$^{\circ}$ phase difference with the transverse displacement \citep{goo14}. Since the Alfv\'en speed increases as the density decreases the further away we are from the tube's axis, but especially across the boundary layer, the periods of oscillation of the azimuthal waves in the resonant layer will be smaller than that of the transverse displacement, leading to a drift in time between the transverse displacement and LOS velocity (a process known as phase mixing). Over long time scales of several periods this process leads to generation of small-scales from which a broadening of the line width, a decrease of the Doppler signal and a loss of coherence is obtained across the resonant layer. In the short observable window of a few periods following an external perturbation (before resonant absorption dampens out the transverse POS motion) a favored 180$^{\circ}$ phase difference stands out. This phenomenon is more clearly seen with a standing wave, for which the phase difference cannot grow indefinitely due to the limited length of the flux tube (Figure \ref{fig6}, see also the schematic representation in Figure \ref{fig7}(a)). 

Apart from resonant absorption, another important mechanism in our simulations is the Kelvin-Helmholtz instability (KHI). As in the coronal case \citep{ter08b,ant14}, the KHI is rapidly triggered at the boundaries after one period of oscillation, leading to the generation of vortices and current sheets (see Figure 2 in Paper 2). Importantly, a complex interplay between resonant absorption and KHI sets in, in which the instability extracts energy from the resonant layer and imparts momentum on the generated eddies, which rapidly degenerate into turbulent-like flows, as can be seen in the online movie B. 

To analyze the overall effect of resonant absorption combined with KHI we performed detailed radiative transfer modeling, explained in the next section. 

\subsection{Radiative transfer}

The chromospheric conditions of our flux tube demand a completely different approach in terms of forward modeling with respect to the previously modeled coronal case \citep{ant14}. In order to properly compare with observations we translate the numerical results into observable quantities by means of appropriate radiative transfer modeling. For optically thin transition region lines such as \ion{Si}{4}~1402.77~\AA\, we calculate the synthetic emission using the FoMo code \citep{ant13}, based on the CHIANTI atomic database version 7 \citep{der97,der09}. For optically thick lines such as \ion{Mg}{2}~h\&k and \ion{Ca}{2}~H\&K we use the RH code, as follows. We employed the two-dimensional version of the RH code \citep[][based on the method described in \citet{ryb92}]{uit01} to calculate the emergent spectra of the \ion{Mg}{2}~h\&k as well as the \ion{Ca}{2}~H\&K resonance lines in non-LTE through a slice perpendicular to the axis of the prominence. The RH code uses the efficient Accelerated Lambda Iteration scheme \citep{ryb92} for transitions with overlapping wavelengths. It uses a formal solution of the transfer equation based on the short-characteristics method \citep{kun88} with Bezier interpolation \citep{aue03}.

Since the MHD code does not provide explicit electron number densities (required to calculate the collisional excitation, de-excitation and ionization-recombination rates), they were computed with the RH code in a two-step process. First, an estimate of electron densities was made under the assumption of LTE. Second, this first estimate was used to calculate a full two-dimensional non-LTE solution of a six-level plus continuum hydrogen atom. Finally, the ionization of hydrogen (i.e., the proton density) gives an estimate of the non-LTE ionization contribution of hydrogen, which was added to the LTE contribution of all other elements (including He).

We used abundances from \citet{gre91} for our calculations, which correspond to photospheric values. For the atomic models we employed the five-level plus continuum model for \ion{Ca}{2} \citep{uit89}, the ten-level plus continuum model for \ion{Mg}{2} \citep{uit97}, and a five-level plus continuum model for Hydrogen with transition values \citep{ver81}, treating all pertinent radiative bound-bound and bound-free transitions explicitly. Since the prominence cross section is optically thick only in the very cores of the lines, we used the approximation of complete frequency redistribution (CRD) in all line transitions, including the Lyman lines in hydrogen. This is justified because wavelengths outside the Doppler core are optically thin and scattering in them does not significantly affect the line transfer in the line cores.

\subsection{RH results}

Through forward modeling of the numerical results we generate observable quantities for direct comparison with the \emph{IRIS} and \emph{Hinode} observations. We first calculate the synthetic emission in \ion{Mg}{2}~h\&k and \ion{Ca}{2}~H\&K with RH on a two-dimensional cross-section at the center of the tube at a given time during the numerical simulation (we pick time $t = 1,162$ s, corresponding to roughly two full periods from the start of the simulation; the numerical model at this time is shown in Figure 2 in Paper 2). In Figure \ref{fig8} we show the source function at the center wavelengths of the \ion{Mg}{2}~k and \ion{Ca}{2}~H lines, as well as the contours for different values of the optical thickness. As can be seen, the source function shows a ring shape with maximum values around the boundary layer of the tube, and negligible values at the center of the tube. This means that the emergent intensity in both of these lines comes from the boundary of such flux tubes.

\section{Discussion}


\subsection{A ring shape for the source function}

The obtained ring shape for the source function in \ion{Mg}{2}~h\&k and \ion{Ca}{2}~H\&K is mostly a temperature effect. Cool prominence plasma around the boundary layer is heated up by the KHI to transition region temperatures, thereby enlarging inwardly the transitional layer from the core to the surrounding corona. The higher temperature in the outer ring raises both the electron density and the electron temperature, both giving rise to more collisional excitations in the h \& k lines (that are then followed by an increased rate of spontaneous emission). The radiative transfer results further suggest that cool chromospheric cores of such flux tubes do not contribute to intensity in \ion{Mg}{2}~h\&k but only to opacity. The opacity in these lines depends on the ground level populations in \ion{Mg}{2} and not on the radiation field. Also, the emerging intensity is optically thin and is therefore mostly dependent on temperature. Accordingly, rays going through the center of the tube are optically thick (in k more than in h, by a factor of 2), so these suffer from self absorption of the signal that comes from the back surface. Such rays have an intensity ratio of k and h lines tending to 1.5. On the other hand, rays going through the surface of the tube only have ratios close to 2 because they are purely optically thin. Ratios between 1.5 and 2 are thus obtained for spatially broad rays (simulating the case of coarse instrument spatial resolution).

An important result from our forward modeling with RH is that the emergent intensity in both the \ion{Mg}{2}~ and \ion{Ca}{2}~H lines comes from the boundary of such flux tubes, where resonant absorption and the KHI vortices take place. Furthermore, the optical thickness values in Figure \ref{fig8} show that the boundary of such flux tubes is optically thin in both lines, while the cool and dense core is significantly opaque only in \ion{Mg}{2} h\&k, an important difference with respect to \ion{Ca}{2} H\&K. This result implies that the former lines suffer less from line-of-sight projection effects than the latter, and are therefore more favorable for prominence analysis at the limb such as that performed here.

This result allows us to perform an approximation to optically thick radiation with FoMo, as explained in Paper~2, which retains the key aspects of the radiative transfer calculations. This approximation allows us to forward model swiftly and thus study in detail the observational signatures of our numerical model. 

\subsection{Interpretation of the observational results: resonant absorption \& KHI}

While the phase drift produced by resonant absorption that results in the transverse MHD wave model (which would correspond to the initial part of phase mixing) could potentially explain the out-of-phase difference found in our observations, it also poses a problem of detectability, and especially it fails to explain the strong dynamical coherence explained in section~\ref{phasedif}. Indeed, resonant absorption is expected to be confined to the fine resonant layer, which is completely invisible to the current instrumentation. Even in inhomogeneous flux tubes with irregular internal density profiles, while the system still evolves with a main oscillating frequency and a large-scale resonance pattern sets in around the ensemble of inhomogeneities, the complicated resonant layers are still found to be highly spatially confined \citep{pas11,ter08c} (this case is sketched in Figure \ref{fig7} (b) and we refer to it as the multiple kink wave scenario). Such systems could potentially explain the POS dynamical coherency (assuming that the inhomogeneities within flux tubes could correspond to the observed threads) but would fail at explaining the LOS velocity coherency and especially the out-of-phase behavior. 

This is where the second mechanism in our simulations, the Kelvin-Helmholtz instability (KHI) becomes important. Indeed, as explained in section~\ref{model} the instability extracts energy from the resonant layer and imparts momentum on the generated eddies, which rapidly degenerate into turbulent-like flows that are still embedded in the large-scale azimuthal flow. This results in a spatial broadening of the transition layer between the prominence core and the external environment, which mostly retains the characteristic phase drift of the resonance layer with respect to the transverse displacement of the prominence core. Furthermore, the layer affected by the KHI, which occupies most of the original flux tube, is subject to significant heating, leading to a fade out of \ion{Mg}{2}~k emission, and gradual appearance of \ion{Si}{4} emission. This is shown in Figure \ref{fig6}, where the Doppler signal captured by the slits (placed a distance of a diameter away from the flux tube, above and below) shows the characteristic out-of-phase behavior with respect to the POS motion of the prominence core. As explained in Paper 2, this is not only the result of the KHI broadening, but also of the PSF width of \emph{IRIS} (which has been taken into account in our forward modelling), which spreads the signal over a significant distance. Furthermore, due to the significantly large perpendicular distance over which the LOS velocity is retained and to the symmetric Doppler profile with respect to the flux tube axis, the dynamical coherence in the observations is recovered. 

As shown in Paper~2, the characteristic observational signatures of the transverse MHD wave model can be seen for basically any LOS in a plane perpendicular to the tube axis, and for optically thin and moderately thick conditions (Paper 2). In the multiple kink wave scenario previously discussed the same observational features would be expected since the ensemble of `threads' (enhanced density regions within the large flux tube) evolve coherently, developing a large azimuthal flow around the system. The onset of the KHI around this flux tube would therefore also produce the same characteristic features as in the present case. 

Another feature predicted by our numerical model is the appearance of thread-like structure in both chromospheric and transition region lines, which result from the KHI vortices and line-of-sight projection effects. Such structure can be seen in Figure \ref{fig6} and is similar to the strand-like structure obtained in the coronal case \citep{ant14}. In Paper~2 we discuss the possibility of the observed \ion{Ca}{2}~H threads being the KHI vortices.

\subsection{Other interpretations}

Our detailed modeling shows that the observations are fully compatible with a state-of-the-art numerical model of transverse MHD oscillations and thus our interpretation of resonant absorption. Furthermore, we can also prove that in the scenario of other possible processes, such as rotation and the torsional Alfv\'en wave, the observed coherence and phase difference between the POS motion and LOS velocity cannot be reproduced.

Let us first consider the case of a torsional Alfv\'en wave. We can further divide this case into 2 different cases that would lead to the observed POS oscillation of threads. Case 1 (sketched in Figure \ref{fig7}(c)) corresponds to a scenario in which the ensemble of threads is subject to an $m = 0$ torsional Alfv\'en wave (as belonging to a larger flux tube). Case 2 (sketched in Figure \ref{fig7}(d)) shows a single thread and corresponds to a scenario in which each thread is subject to a coupled torsional Alfv\'en wave and kink wave, in which the azimuthal flow from the former dominates that of the latter one. In each sketch a generic location for the \emph{IRIS} slits has been taken, in which the width of the slit is set according to the size of the thread, as is observed. Case 1 can be further divided into two cases, depending on the size of the flux tube with respect to the size of the region scanned by the rasters. If the slit positions cover different parts of the flux tube (edges and center, as pictured by case 1 within Figure \ref{fig7}(c)) the LOS velocity signals recovered by the green and red slit positions should be out of phase, disagreeing with the observed transverse coherence in the Doppler signals shown in Figure \ref{fig5} (in each panel of this figure the signal from a slit follows the same behavior regardless of whether it is placed above or below the thread). If the slit positions cover only one half of the flux tube the transverse coherence is recovered (as shown by case 2 in panel (c)), as well as the 180 degree phase difference. However, in this case a gradual increase of amplitude should be observed between the slits the closer they are to the flux tube boundary, which is not observed in our \emph{Hinode}--\emph{IRIS} measurements, as shown in Figure \ref{fig5} in all panels. For the case of coupled torsional Alfv\'en and kink waves the Doppler shift will drastically change sign depending on where the slit is across the thread (as in the first case of the torsional Alfv\'en wave), which impedes the collective in-phase signals between the slits that we observe in Figure \ref{fig5} in all panels (asymmetric $\lambda$--$y$ profile, cf. section 4.3 in Paper 2). This is also shown by \citet{goo14}, in their Figure 9. 

Let us now consider the case of rotation. Similarly, we can identify two cases leading to the observed POS oscillation. The first case (sketched in Figure \ref{fig7}(e)) corresponds to a scenario in which the ensemble of threads is rotating coherently. In the second case (sketched in Figure \ref{fig7}(f)) each thread is subject to rotation coupled to a kink wave, in which the azimuthal flow of the former dominates over that of the latter one. In both cases a slit placed always on one side of the thread should not change sign in time. Also, a slit that crosses the thread should exhibit a change in the sign of the Doppler signal. None of the observed cases in Figure 5 satisfies both of these scenarios, which invalidates rotation as well.

We note that the previous predictions should not be affected by optical thickness. Indeed, for both cases of rotation and a torsional Alfv\'en wave, the frontside and backside of the flux tube with respect to the observer exhibit a coherent motion. Therefore, regardless of how opaque the core of the flux tube is, the signal detected by \emph{IRIS} will retain its Doppler shift value. The predictions should therefore hold, excluding both scenarios of rotation and torsional Alfv\'en wave.

Figure \ref{fig9} summarizes the observed features together with the expected observational features of the 6 considered cases. The plus/minus sign denotes a change in the LOS velocity. The green and red curves denote the shape of the LOS velocity in time, as detected by the slits of same color in the sketches.

\section{Conclusions}

Observations of coronal transverse oscillations were first (and are often) reported in the aftermath of flares \citep{nak99}. Such reports have been the subject of several dozen follow-up theoretical and observational studies, which suggest resonant absorption as the most plausible mechanism causing the damping of such oscillations \citep{dem12,nak05}. However, to the date no solid consensus exists on the causes of this damping. Furthermore these large-scale oscillations occur only in the aftermath of an impulsive event such as a flare. Such events cannot be the main mechanism of heating in the quiescent corona as they are too infrequent (compared to the typical coronal cooling times). Heating from Alfv\'enic waves invoking resonant absorption in prominences or coronal loops has also previously been reported \citep{nak99,ofm96,ofm98}, supported by numerous theoretical studies suggesting that this mechanism is efficient and ubiquitous \citep{arr08,goo11,ter08c}. However, such observational reports have been largely based on indirect evidence due to incomplete measurements (lack of spectroscopic information allowing the detection of the three-dimensional flow and heating generated by the wave) or lack of resolution leading to a precarious causal relation between resonant absorption and the subsequent heating. In the present work we show for the first time direct evidence of resonant absorption as measured with a spectrograph and associated heating to transition region temperatures. This association is performed with unprecedented detail and therefore narrows down to unparalleled levels the causal relation in evidence for wave heating.

\begin{figure}
\epsscale{1}
 \plotone{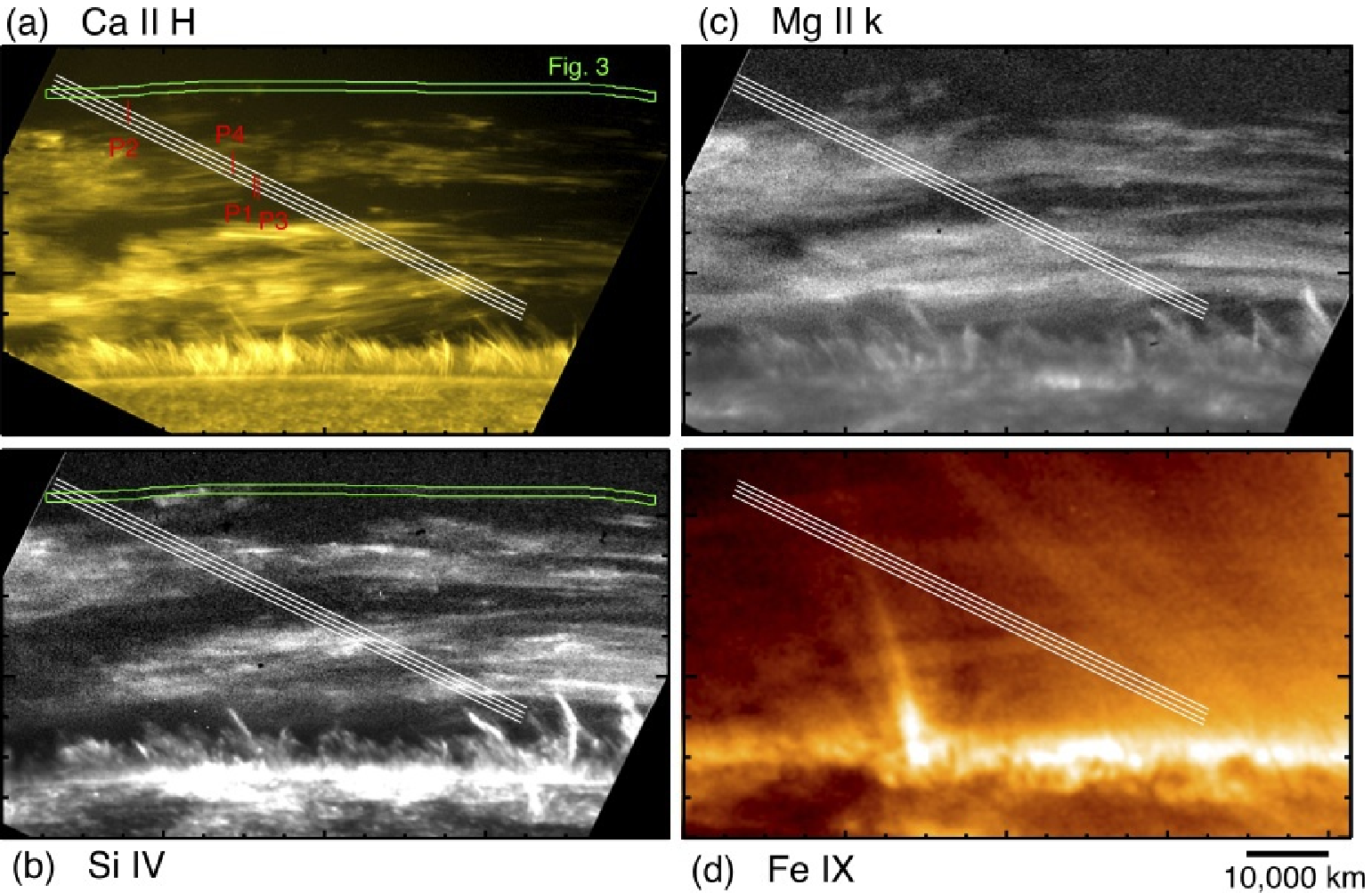}
    \caption{
Prominence observed on the southeast solar limb on 19 October 2013. (a) \emph{Hinode}/SOT image of the observed prominence in the \ion{Ca}{2} H line ($\sim$10,000 K). The horizontal elongated structures are prominence threads. The vertical ones are spicules, which are not studied here. The four white lines are the positions of the \emph{IRIS} slit. The green box indicates the region used for tracking the horizontally-moving threads for Figure \ref{fig3}. The vertical red bars are the selected locations for the space-time plots in Figure \ref{fig4}. (b) \emph{IRIS} slit-jaw image in the \ion{Si}{4} line ($\sim$80,000 K). The image is processed to remove the dark slit with partial substitution of the previous-time image. (c) \emph{IRIS} slit-jaw image in the \ion{Mg}{2} line (10,000--15,000 K). The same process was applied as for the \ion{Si}{4} image. (d) \emph{SDO}/AIA image in the \ion{Fe}{9} passband image ($\sim$1 MK). An active region (AR 11877) is located on the right-hand side out of the field of view (FOV). Bright loops that are superposed in front of the prominence along the LOS come from the edge of the active region. See the online movie A for time series of these images.
(An animation of this figure is available.)
}
    \label{fig1}
\end{figure}

\begin{figure}
\epsscale{1.0}
 \plotone{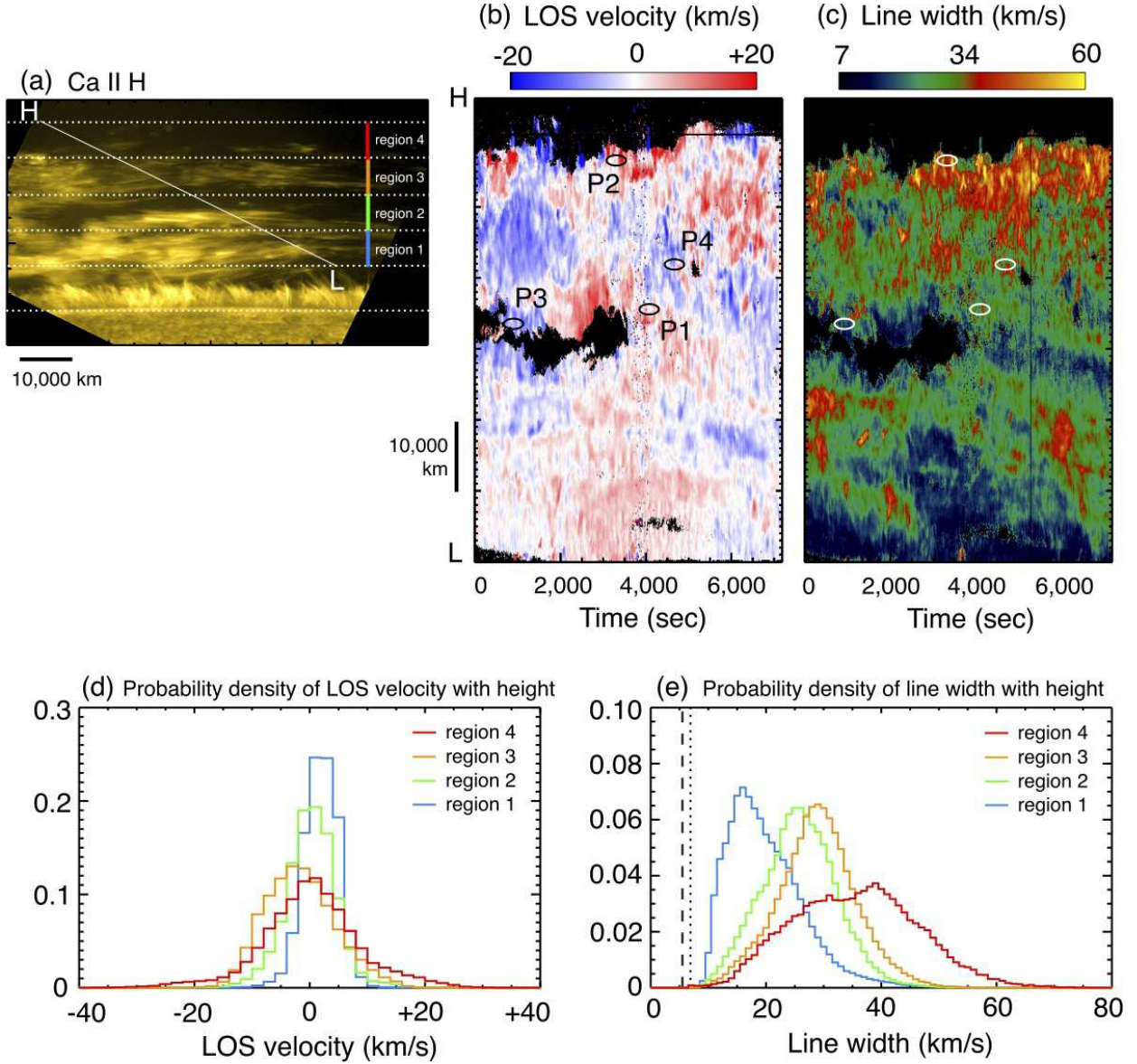}
    \caption{
Time variation of LOS velocity and line width and height dependence of oscillations. (a) \emph{Hinode}/SOT image in the \ion{Ca}{2} H line. The solid white line is the location of the \emph{IRIS} slit (the second slit from the top shown in Figure \ref{fig1}). (b) Space-time plot of LOS velocity derived from \ion{Mg}{2} k spectra along the slit shown on panel (a). The notations L and H on the left axis correspond to the height locations indicated by the same ones on panel (a). (c) The same as panel (b), but of line width. Note that the values on panel (c) are the full width at half-maximum (FWHM). (d--e) Height dependence of LOS velocity and line width binned for different height ranges separated by 10\arcsec\ (7,200 km on the Sun) indicated on panel (a). The dashed and dotted lines on panel (e) indicate the instrumental line broadening (5.5 km s$^{-1}$ in FWHM) and combination of the instrumental and thermal broadenings (6.9 km s$^{-1}$ in FWHM), respectively.
}
    \label{fig2}
\end{figure}

\begin{figure}
\epsscale{1.0}
 \plotone{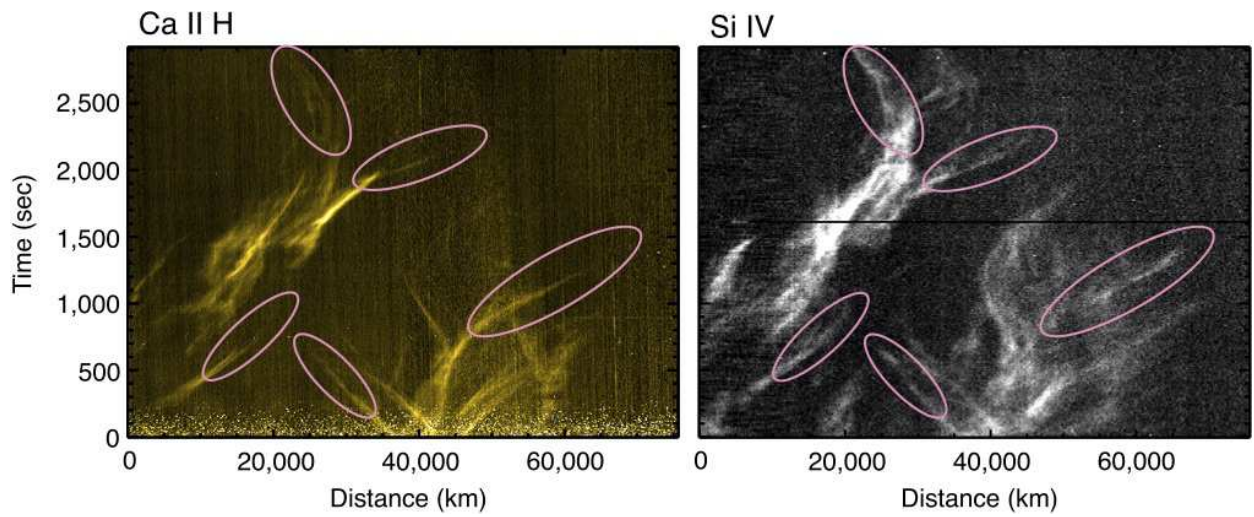}
    \caption{
Heating of prominence threads. Left: Space-time plot of \emph{Hinode}/SOT \ion{Ca}{2} image ($\sim$10,000 K) along the green box shown in Figure \ref{fig1}. The intensity in the green box is summed vertically to include the entire part of oscillating threads. Right: The same as the left panel, but for \emph{IRIS} \ion{Si}{4} image ($\sim$80,000 K). Several paths (indicated by circles) show the transition of temperature from cool to hot, as the threads in the \ion{Ca}{2} line fade away and co-spatial threads appear in the \ion{Si}{4} line.
}
    \label{fig3}
\end{figure}

\begin{figure*}
\epsscale{1.0}
 \plotone{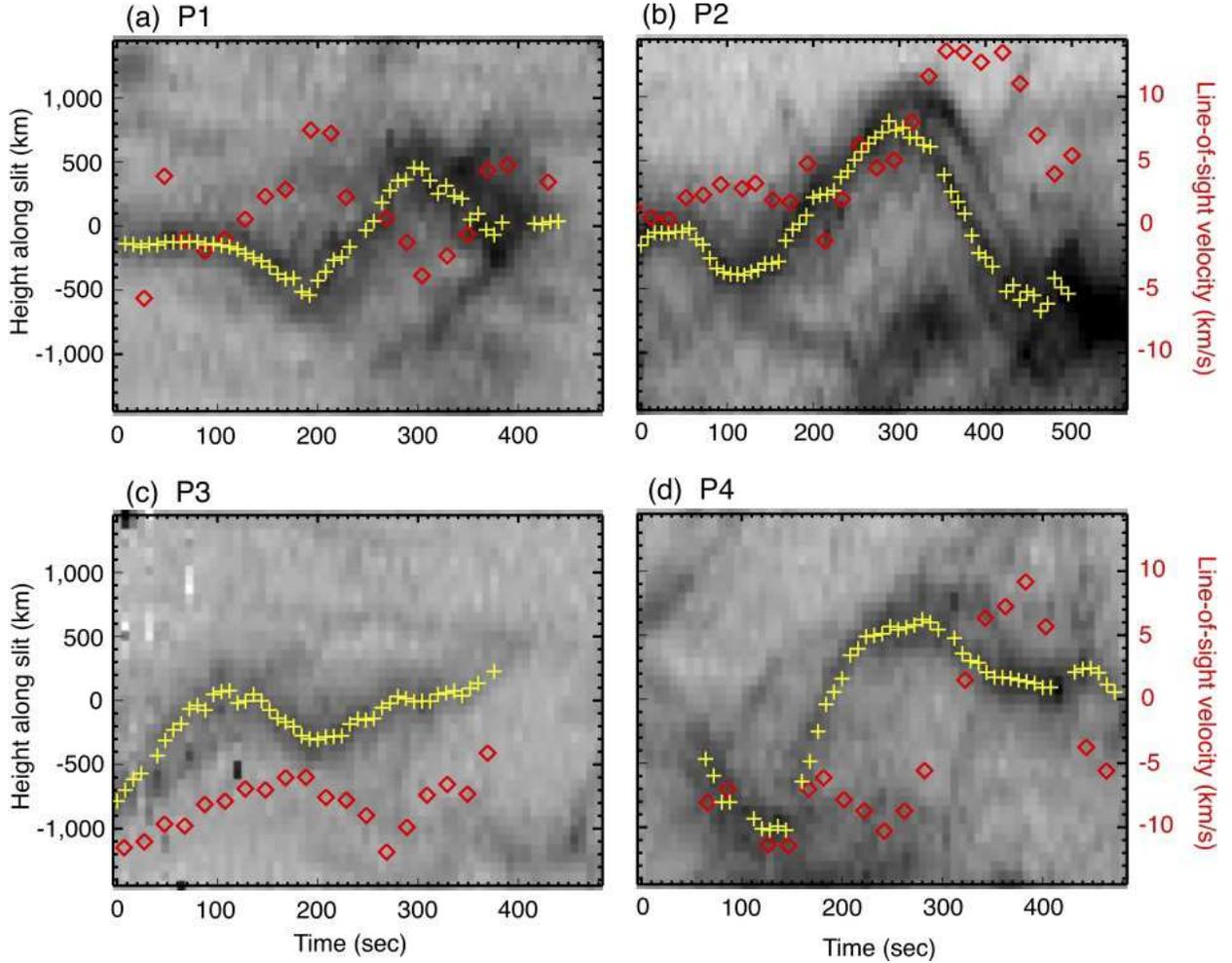}
    \caption{
Examining phase difference between transverse displacement and LOS velocities. Space-time plots of \emph{Hinode}/SOT Ca images at four locations shown by the red bars (P1--P4) in Figure \ref{fig1}. The yellow crosses are the central positions of the threads. The red diamonds indicate the LOS velocity derived from \ion{Mg}{2} k spectra at the corresponding positions of the thread center. (a) P1 shows clear oscillations with 180$^{\circ}$ phase difference between the transverse motions and LOS velocities. (b) P2 presents similar, but with 90$^{\circ}$ difference. (c--d) P3 and P4 show phase difference between 90$^{\circ}$ and 180$^{\circ}$.
}
    \label{fig4}
\end{figure*}

\begin{figure*}
\epsscale{1.0}
 \plotone{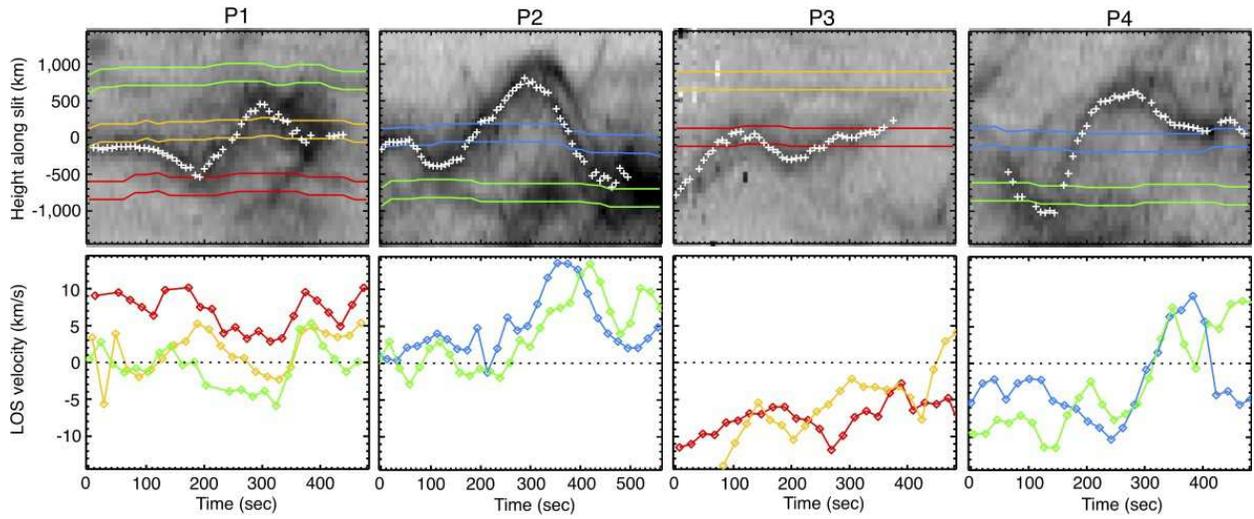}
    \caption{
Phase difference between transverse displacement and LOS velocities at neighboring slit locations. Similar to Figure \ref{fig4}, but for all LOS velocity information at and near the oscillating threads. The slit locations are shown by colored lines on the upper panels. The colored diamonds are the LOS velocities along the respective colored slits. We can find some coherence of the velocity patterns over a significant distance away from the threads.
}
    \label{fig5}
\end{figure*}

\begin{figure*}
\epsscale{1.0}
 \plotone{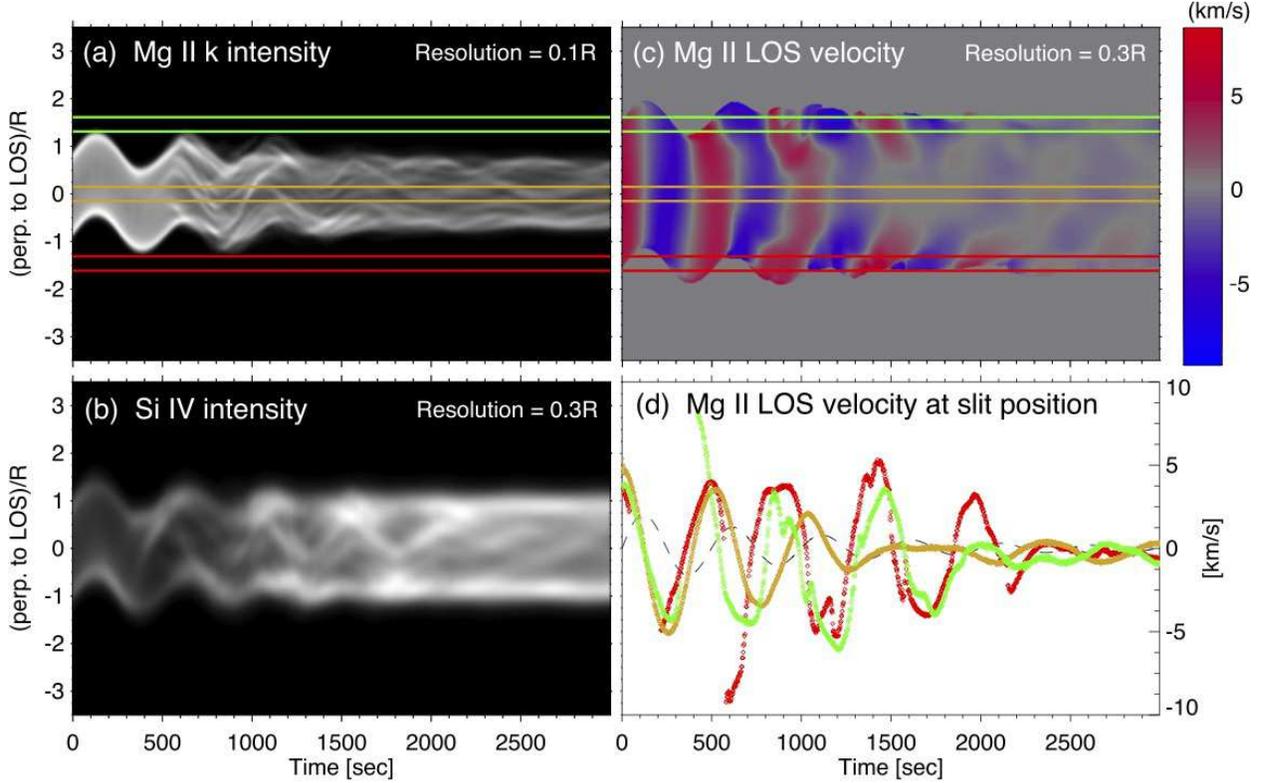}
    \caption{
Synthetic intensity and LOS velocity from a simulated prominence flux tube with a transverse oscillation for a LOS angle of 45$^{\circ}$ in the plane perpendicular to the tube axis. (a) Intensity of the \ion{Mg}{2} k line with a spatial resolution of $0.1 R$ ($R$ being the radius of the tube). (b) Intensity of the \ion{Si}{4} line with a spatial resolution of $0.3 R$. (c) Space-time plot of the LOS velocity for a slit perpendicular to the tube axis located at the center of the tube with a spatial resolution of $0.3 R$. An arrow shaped pattern develops due to resonant absorption. (d) Doppler signals along the green, orange and red slit locations, with respective colors (whose colored edges are shown in panels (a) and (c)), centered at a distance of $1.5 R$, $0 R$ and $-1.5 R$ from the tube's center, respectively. The black dashed line corresponds to the transverse displacement of the tube in the POS, calculated from Gaussian fits to the intensity image (a) for each time step. Times without Doppler signal correspond to regions for which the signal is too low. The phase differences between the LOS velocities and POS displacements match those in the observations (Figure \ref{fig4}).
}
    \label{fig6}
\end{figure*}

Our numerical modeling suggests that an important coupling between resonant absorption and KHI takes place, in which KHI efficiently extracts energy from the resonant layer and converts it into heat. As the KHI sets in viscous and ohmic heating strongly increase due to the generation of vortices and current sheets around most of the flux tube \citep{goo02}. As explained previously, the heating occurs rapidly at first and proceeds to a roughly constant rate for most of the simulation. The boundary layer, characterized by transition region temperatures around $10^{5}$ K, expands rapidly in the timescale of one period, mostly inwards towards the center of the tube, which leads to fading with time of threads in \ion{Ca}{2} and \ion{Mg}{2} (both 10,000 K), and the subsequent appearance (in the same location) of threads at higher temperatures on a timescale of 10--15 min (e.g., \ion{Si}{4} at 80,000 K) (Figure \ref{fig6}). Such heating is also present in our observations on similar timescales: as the moving threads in the \ion{Ca}{2} line fade away (the online movie A), co-spatial threads appear in the hotter \ion{Si}{4} line with similar horizontal speeds. Space-time plots of the \ion{Ca}{2} and \ion{Si}{4} images show this clearly (Figure \ref{fig3}). Even though some locations show coexistence of cooler and hotter materials, the trajectories of several threads show a transition from lower to higher temperatures with threads finally fading out even from \ion{Si}{4} images. This scenario cannot be explained by a multi-thermal structure of the threads, and is strongly suggestive of cool plasma being heated to coronal temperatures through the mid-temperature range of the \ion{Si}{4} line. Evidence for coronal counterparts of the prominence comes from the online movie A: while clear fine structures cannot be identified in the \ion{Fe}{9} ($\sim$1 MK) channel of \emph{SDO}/AIA, horizontally-elongated bright features do occupy the same region as in the cooler passbands (Figure \ref{fig1}).

\begin{figure}
\epsscale{1.0}
 \plotone{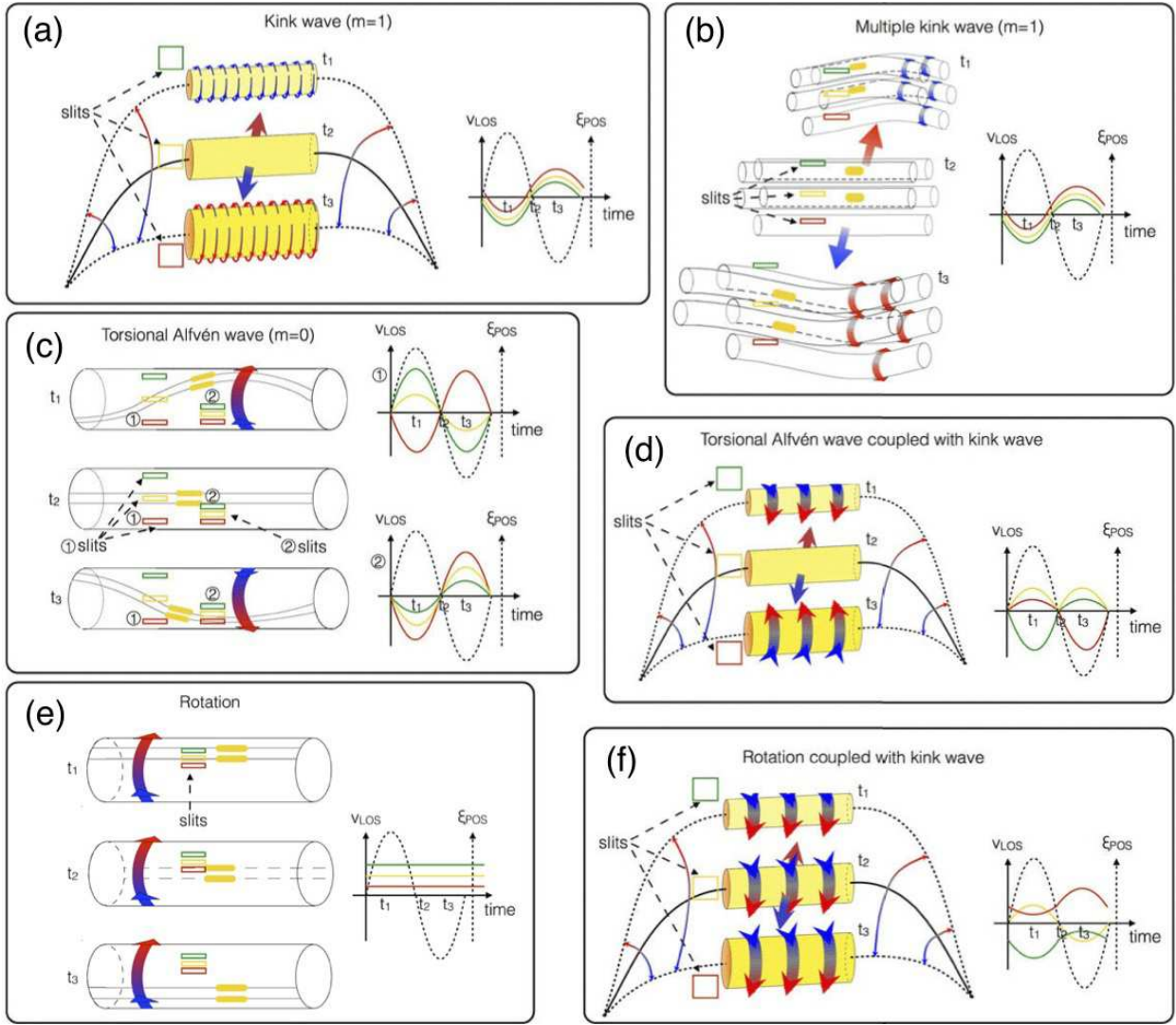}
    \caption{
Schematic representation of the physical model explaining our observations and other considered interpretations. (a-b) The transverse MHD wave model: a single flux tube and multiple flux tubes subject to a kink wave. (c-d) Flux tube with a torsional Alfv?n wave alone or coupled with a kink wave. (e-f) Flux tube with rotation alone or coupled with a kink wave. The observed threads can each correspond to a flux tube (a,d,f) or to high density regions within a flux tube (b,c,e). Locations of the \emph{IRIS} slit are sketched as green, yellow and red boxes, as in the observations (\ref{fig5}). The width of the slit corresponds to that of the thread. The predicted Doppler signal at each slit location (solid) and the POS motion (dashed) in time is sketched on the right side in each panel. For (c), two cases can be distinguished depending on the location of the slits over the flux tube. In each case the middle position is the equilibrium position for an oscillation. In the top/bottom position, the thread is the furthest away/closest from/to the observer. For the transverse MHD wave model (a,b) the azimuthal motions result in a blue/red shift at the topmost/bottom position (blue/red arrows), as observed. None of the other scenarios match the observations.
}
    \label{fig7}
\end{figure}

\begin{figure}
\epsscale{1.0}
 \plotone{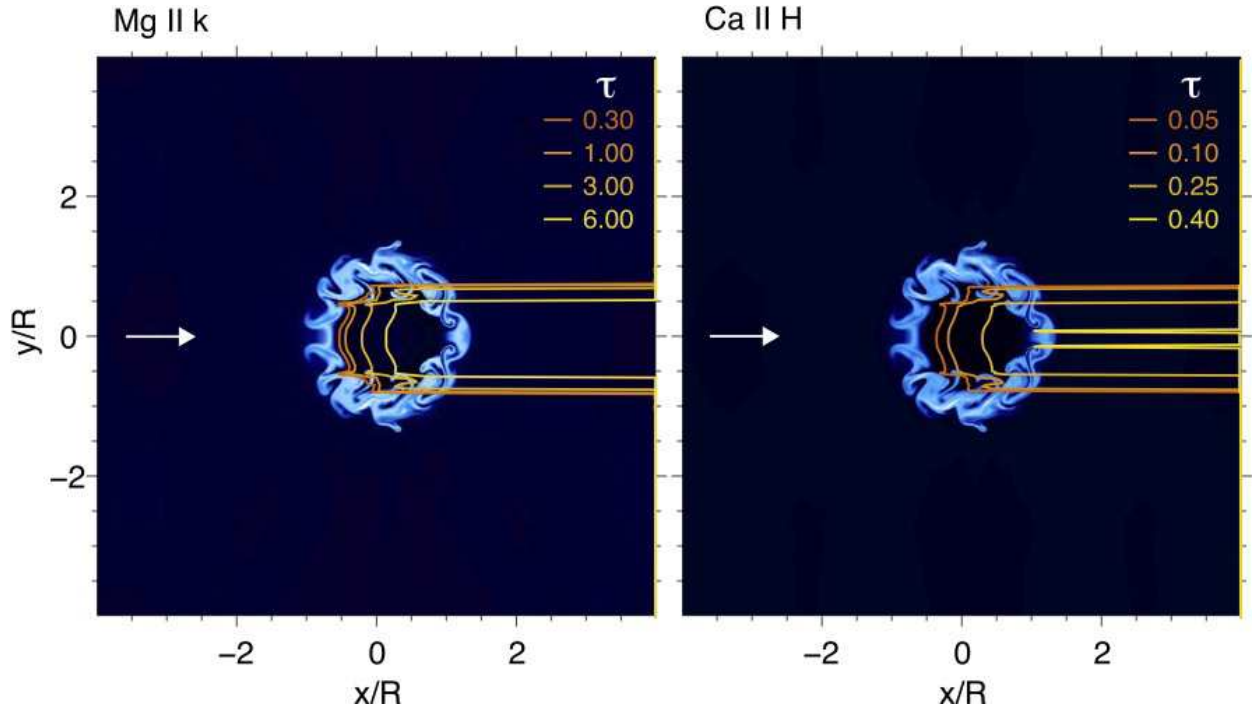}
    \caption{
Source function and optical thickness in \ion{Mg}{2} k and \ion{Ca}{2} H lines. Source function in \ion{Mg}{2} k 2796.35 \AA\ (left) and \ion{Ca}{2} H line center (right) for a cross-section of the tube along the tube's center. The yellow curves correspond to contour lines for different values of the optical thickness (stated in the legends) as seen from the top (as indicated by the arrow), corresponding to an LOS angle of 0$^{\circ}$. The axes are set to match those in the numerical simulation in Paper 2.
}
    \label{fig8}
\end{figure}

\begin{figure}
\epsscale{1.0}
 \plotone{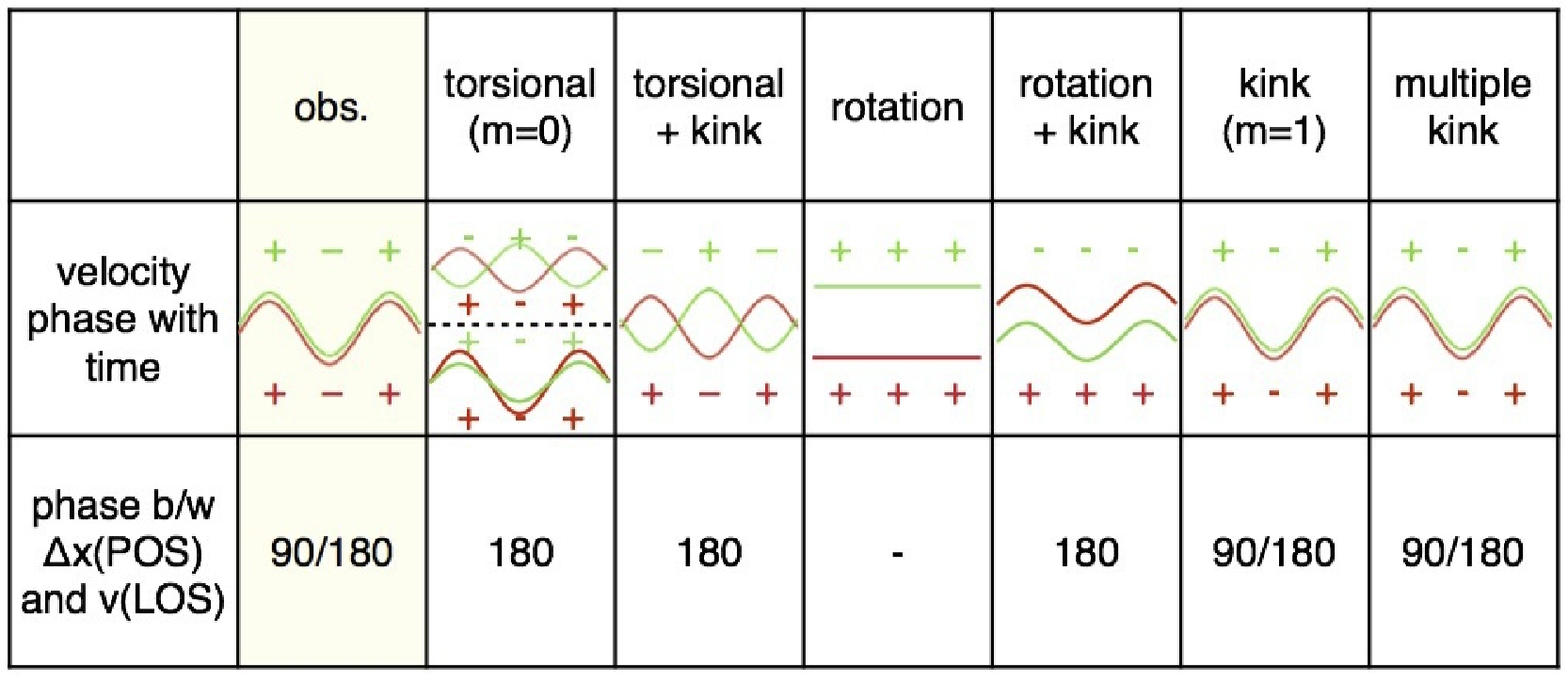}
    \caption{
Summary of all the physical scenarios considered. Stating the expected LOS velocity phase with time (together with the changes of sign, denoted as plus or minus sign) at the two green and red slit locations considered in the previous sketches, and the expected phase difference between the POS motion of the thread and the LOS velocity. $x/y$ notation indicates phase differences of $x$ to $y$. The first column denotes the observations. Our numerical modeling applies to the last two columns. These are the only cases among those considered above which correctly match the observed features.
}
    \label{fig9}
\end{figure}

The observed heating is directly predicted by our transverse MHD wave model and thus provides supporting evidence for our interpretation, which, as mentioned, is focused on the spatial coherence and the phase relationships between LOS velocity and POS oscillations. The combination of observations and modeling thus provide compelling evidence for wave dissipation through the combination of resonant absorption and KHI and resulting heating to at least transition region temperatures in a prominence. The observed oscillations have an energy flux of 100--1,000 W m$^{-2}$, where the velocity amplitude is 20 km s$^{-1}$ in the case that the view angle is 45$^{\circ}$, the plasma density is typically 10$^{-11}$--10$^{-10}$ kg m$^{-3}$ \citep{lab10}, and the Alfv\'en speed is assumed to be 300--600 km s$^{-1}$. Here we have considered the more general case of the multiple kink wave scenario (sketched in Figure \ref{fig7}(b)), for which the energy flux is within a factor of 2 from that of the bulk Alfv\'en wave multiplied by a filling factor \citep{goo13,van14}. We take a conservative filling factor of 5\% to avoid overestimation. The energy flux is enough to locally heat the plasma to coronal temperatures \citep{wit77}. Our results may also provide insight into the puzzling rapid disappearance of prominences \citep{tan95} and in the heating mechanism of hot plasma bubbles that have been reported in prominences \citep{ber11}. More generally, our results provide a pathfinder to search for heating resulting from resonant absorption in other solar regions (e.g., quiet corona and coronal holes) that are permeated with transverse magnetic waves.

\acknowledgements 

\emph{Hinode} is a Japanese mission developed and launched by ISAS/JAXA, with NAOJ as domestic partner and NASA and STFC (UK) as international partners. It is operated by these agencies in cooperation with ESA and NSC (Norway). \emph{IRIS} is a NASA small explorer mission developed and operated by LMSAL with mission operations executed at NASA Ames Research center and major contributions to downlink communications funded by NSC through an ESA PRODEX contract. \emph{SDO} is part of NASA's Living With a Star Program. Numerical computations were carried out on Cray XC30 at the Center for Computational Astrophysics, NAOJ. T.J.O thanks Ippon-kakou-kai for their encouragement and was supported by JSPS KAKENHI Grant Number 25800120 (PI: T.J.O.). T.J.O and P.A. were supported by JSPS KAKENHI Grant Number 25220703 (PI: S. Tsuneta). B.D.P. was supported by NASA under contract NNG09FA40C (\emph{IRIS}), NNX11AN98G and NNM12AB40P. T.V.D. was supported by FWO Vlaanderen's Odysseus programme, GOA-2015-014 (KU Leuven) and the IAP P7/08 CHARM (Belspo).




\end{document}